\documentclass[prl, amsmath,superscriptaddress, floatfix,byrevtev]{revtex4}

\usepackage[dvips]{graphicx}

\begin{document}

\title{X-ray Astronomy in the Laboratory with a Miniature Compact Object Produced by Laser-Driven Implosion}

\author{Shinsuke Fujioka}
\email{sfujioka@ile.osaka-u.ac.jp}
\affiliation{Institute of Laser Engineering, Osaka University, 2-6 Yamada-Oka, Suita, Osaka, 565-0871 Japan}
\author{Hideaki Takabe}
\affiliation{Institute of Laser Engineering, Osaka University, 2-6 Yamada-Oka, Suita, Osaka, 565-0871 Japan}
\author{Norimasa Yamamoto}
\affiliation{Institute of Laser Engineering, Osaka University, 2-6 Yamada-Oka, Suita, Osaka, 565-0871 Japan}
\author{David Salzmann}
\affiliation{Institute of Laser Engineering, Osaka University, 2-6 Yamada-Oka, Suita, Osaka, 565-0871 Japan}
\author{Feilu Wang}
\affiliation{National Astronomical Observatories, Chinese Academy of Sciences, Beijing 100012, China}
\author{Hiroaki Nishimura}
\affiliation{Institute of Laser Engineering, Osaka University, 2-6 Yamada-Oka, Suita, Osaka, 565-0871 Japan}
\author{Yutong Li}
\affiliation{Beijing National Laboratory for Condensed Matter Physics, Institute of Physics, Chinese Academy of Sciences, Beijing 1000190, China.}
\author{Quanli Dong}
\affiliation{Beijing National Laboratory for Condensed Matter Physics, Institute of Physics, Chinese Academy of Sciences, Beijing 1000190, China.}
\author{Shoujun Wang}
\affiliation{Beijing National Laboratory for Condensed Matter Physics, Institute of Physics, Chinese Academy of Sciences, Beijing 1000190, China.}
\author{Yi Zhang}
\affiliation{Beijing National Laboratory for Condensed Matter Physics, Institute of Physics, Chinese Academy of Sciences, Beijing 1000190, China.}
\author{Yong-Joo Rhee}
\affiliation{Laboratory for Quantum Optics, Korea Atomic Energy Research Institute, 1045 Daedeok Street Yuseong-gu, Daejon 305-353, Korea.}
\author{Yong-Woo Lee}
\affiliation{Laboratory for Quantum Optics, Korea Atomic Energy Research Institute, 1045 Daedeok Street Yuseong-gu, Daejon 305-353, Korea.}
\author{Jae-Min Han}
\affiliation{Laboratory for Quantum Optics, Korea Atomic Energy Research Institute, 1045 Daedeok Street Yuseong-gu, Daejon 305-353, Korea.}
\author{Minoru Tanabe}
\affiliation{Institute of Laser Engineering, Osaka University, 2-6 Yamada-Oka, Suita, Osaka, 565-0871 Japan}
\author{Takashi Fujiwara}
\affiliation{Institute of Laser Engineering, Osaka University, 2-6 Yamada-Oka, Suita, Osaka, 565-0871 Japan}
\author{Yuto Nakabayashi}
\affiliation{Institute of Laser Engineering, Osaka University, 2-6 Yamada-Oka, Suita, Osaka, 565-0871 Japan}
\author{Gang Zhao}
\affiliation{National Astronomical Observatories, Chinese Academy of Sciences, Beijing 100012, China}
\author{Jie Zhang}
\affiliation{Beijing National Laboratory for Condensed Matter Physics, Institute of Physics, Chinese Academy of Sciences, Beijing 1000190, China.}
\affiliation{Department of Physics, Shanghai Jiao Tong University, Shanghai 200240, China.}
\author{Kunioki Mima}
\affiliation{Institute of Laser Engineering, Osaka University, 2-6 Yamada-Oka, Suita, Osaka, 565-0871 Japan}

\begin{abstract}
Laboratory spectroscopy of non-thermal equilibrium plasmas photoionized by intense radiation is a key to understanding compact objects, such as black holes, based on astronomical observations \cite{Tarter_69_ApJ, Osterbrock_text, Liedahl_99_text, Ballantyne_02_MNRAS}.
This paper describes an experiment to study photoionizing plasmas in laboratory under well-defined and genuine conditions.
Photoionized plasma is here generated using a 0.5-keV Planckian x-ray source created by means of a laser-driven implosion \cite{ICF_concept}.
The measured x-ray spectrum from the photoionized silicon plasma resembles those observed from the binary stars Cygnus X-3 \cite{Paerels_00_ApJ, Liedahl_96_ApJ} and Vela X-1 \cite{Schulz_02_ApJ, Goldstein_04_AJ, Watanabe_06_ApJ} with the Chandra x-ray satellite.
This demonstrates that an extreme radiation field was produced in the laboratory, however, the theoretical interpretation of the laboratory spectrum significantly contradicts the generally accepted explanations in x-ray astronomy.
This model experiment offers a novel test bed for validation and verification of computational codes used in x-ray astronomy.
\end{abstract}

\maketitle

	X-ray spectroscopy with x-ray satellite is the main observational method to give information about compact objects, especially black holes (BHs).
	BHs are indirectly studied by observing the x-ray continuum from a heated accretion disk and x-ray fluorescence from the ambient gas of the stellar wind and the surface of a companion star in their binary systems.
	To derive physical properties from the observations, x-ray astronomers rely on non local-thermal-equilibrium (LTE) atomic physics in a cold ambient gas subject to an extreme radiation field, whose mean radiation temperature is on the order of 1 keV.
	Theoretical models have been developed on the basis of the observed spectra \cite{Tarter_69_ApJ, Osterbrock_text, Liedahl_99_text, Ballantyne_02_MNRAS} and complex computer codes were developed to analyze the observational x-ray spectra \cite{Kallman_96_ApJ, Ferland_98_PASP, Rose_98_JPB, Chung_05_HEDP, Rose_04_JPB}.
	The underlying assumption of these models is that the spectrum originates from a \textit{photoionized plasma}. In other words, the intense radiation from the compact object photoionizes the gas, and generates a relatively low electron-temperature highly-ionized non-LTE plasma. 
	However, laboratory experiments on non-LTE photoionized plasmas have not been available, mainly owing to the lack of an intense source of x-ray continuum radiation.
	Only recently did pulsed power apparatus, laser and $Z$-pinch, reproduce the extreme conditions in the universe \cite{Remington_99_Science, Remington_06_RMP, Foord_04_PRL, Wang_08_PoP}.
	
\begin{figure}
	\begin{center}
		\includegraphics*[width= 7cm]{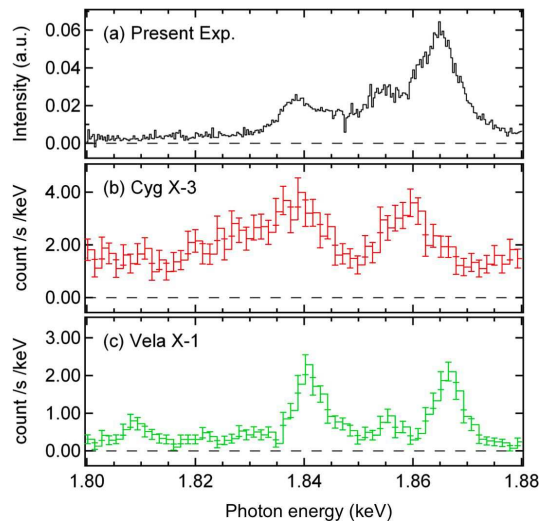}
	\end{center}
	\caption{Comparison of x-ray spectra for (a) a photoionized plasma generated in the laboratory and those observed from astronomical objects, (b) Cygnus X-3 and (c) Vela X-1. The emission peaks near 1.86 - 1.87, 1.85 - 1.86, and 1.84 keV are identified in astronomy to stem from resonance, intercombination, and forbidden transitions of He-like silicon ions, respectively. \label{fig: spectra}}
\end{figure}

	Here we present the terrestrial experiment of non-LTE photoionized plasmas.
    The novelty of the present experiment is the notion that laser-driven implosion can create a flash of brilliant Planckian x-ray source that can be used to simulate a miniature astronomical compact object.
	X-ray spectra having two characteristic spectral peaks were observed for a photoionized silicon plasma generated in the laboratory.
	This spectral shape resembles closely those observed from Cygnus X-3 and Vela X-1, as shown in Figs. \ref{fig: spectra} (a) - (c).
	In Figs. \ref{fig: spectra} (a) and (c), even the small bump between the two peaks is reproduced.
		
\begin{figure}
	\begin{center}
		\includegraphics*[width= 7cm]{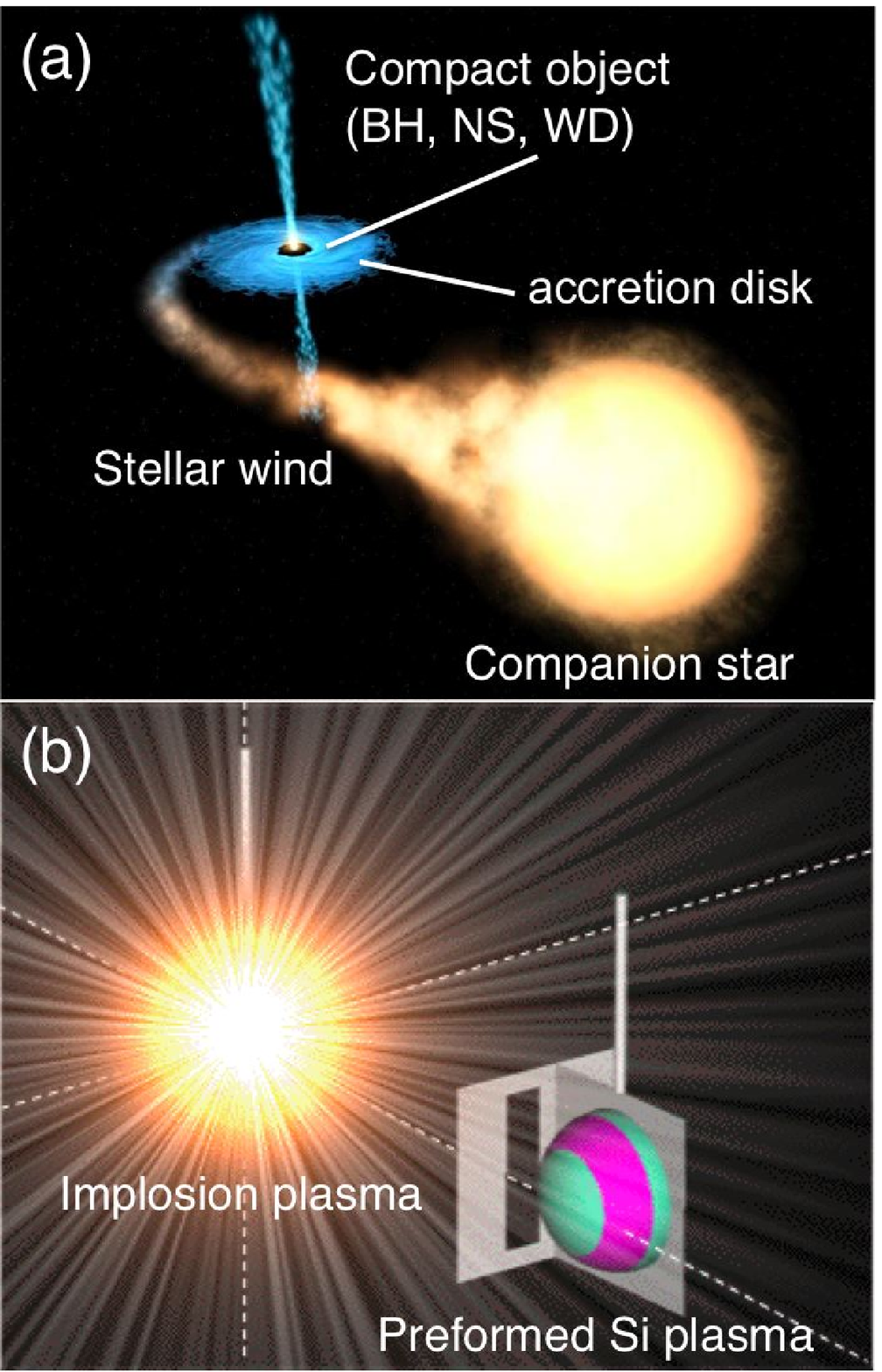}
	\end{center}
	\caption{(a) Artist conception of a binary system consisting of a black hole and a companion star. Ambient gases are photoionized by strong radiation emitted from the accretion disk. [These images were created for the European Space Agency by the Hubble European Space Agency Information Centre and for NASA by STScI under Contract NAS5-26555.] (b) Schematic view of the photoionized plasma experiment. A spherical hollow plastic shell is imploded with twelve laser beams from the GEKKO-XII facility. The resulting core plasma mimics a compact object, a Planckian x-ray radiator with a radiation temperature of 480 $\pm$ 20 eV. Silicon plasma with a 30-eV temperature was produced in the vicinity of the Planckian radiator. A tantalum plate, with a slit in it, was inserted between the silicon plasma and the x-ray source to prevent the direct illumination of x-ray from the imploded core onto the silicon surface. \label{fig: schematic}}
\end{figure}

	Cygnus X-3 is a well known x-ray object identified in the early stages of x-ray astronomy \cite{Giacconi_67_APL}.
	It is a binary system consisting of a BH candidate and a companion star. 
	An artist conception of such a binary system is shown in Fig. \ref{fig: schematic} (a), in which the gravitational energy of the accreting material is converted into thermal energy, which is the origin of the strong radiation emitted from the accretion disk \cite{Shakura_73_AA}.
	Figure \ref{fig: spectra} (b) shows an x-ray spectrum from Cygnus X-3 observed with a spectrometre onboard the \textit{Chandra} X-ray satellite.
		The spectrum is redshifted by 800 km/s \cite{Paerels_00_ApJ}.
	Line x-rays from highly ionized silicon ions are emitted from the surface of the companion star, whose area is much larger than that of the accretion disk and the BH.
	The electron temperature of the surface is determined to lie in the range of 5 - 50 eV by fitting the spectral shape of several radiative recombination continua \cite{Liedahl_96_ApJ, Paerels_00_ApJ}.
	This temperature is too low to ionize silicon atoms to H- and He-like ions.
	This fact is direct evidence that the lines in the keV range are due to photoionization by high-energy photons around the BH.
	Similar x-ray spectra were also observed from Vela X-1, a neutron star binary system\cite{Schulz_02_ApJ, Goldstein_04_AJ, Watanabe_06_ApJ} as graphed in Fig. \ref{fig: spectra} (c).
	
\begin{figure}
	\begin{center}
		\includegraphics*[width= 7cm]{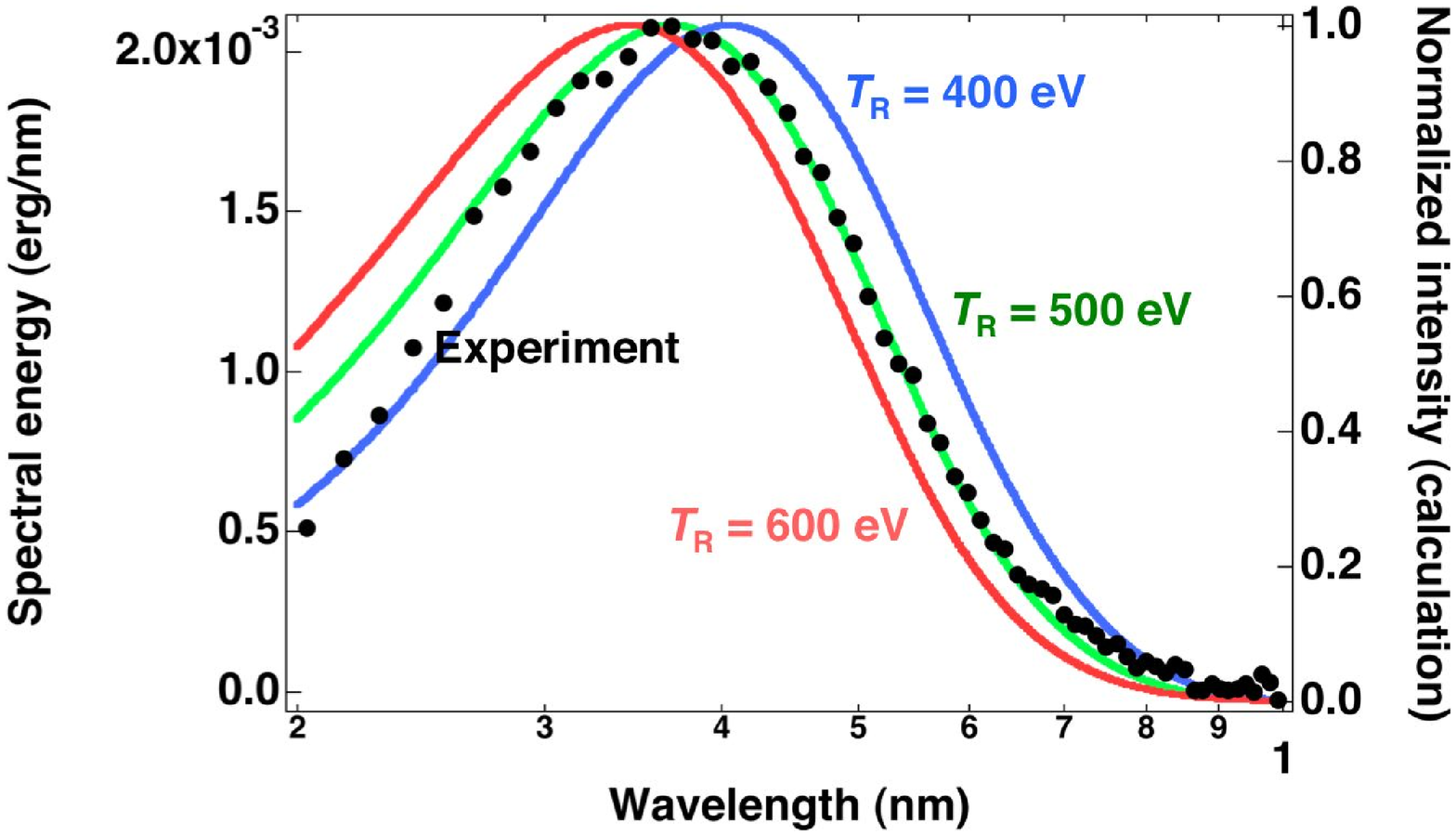}
	\end{center}
	\caption{X-ray spectrum from the core plasma measured with a transmission grating spectrometre. The Planckian spectrum for 500 eV of radiation temperature shows the best agreement with the experiment after accounting for the spectral response of the spectrometre.	\label{fig: planckian}}
\end{figure}

	A direct laser-driven implosion \cite{ICF_concept} was used to create a hot, dense plasma.
	A spherical hollow plastic shell, whose diametre and thickness were respectively 500 $\mu$m and 6 $\mu$m, was imploded by twelve beams from the GEKKO-XII laser facility \cite{Yamanaka_81_IEEEQE} carrying 6 kJ of total energy as 1.2-ns green ($\lambda_L$ = 0.53 $\mu$m) laser pulses.
	The imploded core approaches 1 keV in temperature, and its areal density (column density) reaches 0.1 g/cm$^2$, high enough to be optically thick for a few keV x-rays.
	The measured x-ray spectrum from the imploded core plasma was compared with calculated Planckian spectra modified by the spectral response of the spectrometre consisting of a transmission grating, a CCD camera, and several x-ray filters.
	The resulting spectrum for a 500-eV blackbody exhibits the best agreement with the measured one, as shown in Fig. \ref{fig: planckian}.
	A separate measurement consistently indicates 480 $\pm$ 20 eV as the radiation temperature.
	The duration of the radiation pulse was measured with an x-ray streak camera to be 160 ps, which is long enough to study atomic processes of interest as discussed below.
	The silicon foil was heated by a second, weak infrared ($\lambda_L$ = 1.064 $\mu$m) laser pulse, whose intensity and duration were respectively 5 $\times$ 10$^{10}$ W/cm$^2$ and 10 ns. Consequently, a slowly expanding, low density silicon cloud mimicking ambient gas is produced.
		The electron temperature and density of the silicon plasma were 26 - 29 eV and (0.5 - 1.0) $\times$ 10$^{20}$ cm$^{-3}$, respectively.
		The radiation from the imploded core irradiates part of the silicon plasma through the slit, as shown in Fig. \ref{fig: schematic} (b).
		The silicon plasma was located 1.2 mm from the imploded core and the dilution factor of the Planckian radiation was (3 - 10) $\times$ 10$^{-4}$ at the silicon foil.
			 A 500-eV Planckian radiation source and a silicon plasma were generated simultaneously, only then x-ray lines from He-like silicon ions were clearly measured, as shown in Fig. \ref{fig: spectra} (a).
	The spectral resolution in the experiment was $\Delta h\nu$ = 7 eV due to the size of the photoionized plasma (500 $\times$ 500 $\mu$m$^2$).
	This resolution is similar to that of the astronomical observations in Figs. \ref{fig: spectra} (b) and (c).
	
	  In the astrophysical literature \cite{Tarter_69_ApJ, Foord_04_PRL}, the ionization parameter is defined as $\xi = 16 \pi^2 J/n_{e}$ (with units of erg cm/s)  to measure importance of photoionization in a plasma, here $J$ is the mean radiation intensity per steradian, integrated over solid angle and integrated in photon energy.
In the region of photoionized plasma, highly ionized ions are observed by UV and x-ray satellites over the range $\xi$ = 10 - 10$^4$ ergs cm/s. In the present experiment, the ionization parameter is $\xi$ = 5.9 $\pm$ 3.8, this is only slightly below the astrophysical value.
	
\begin{figure}
	\begin{center}
		\includegraphics*[width= 7cm]{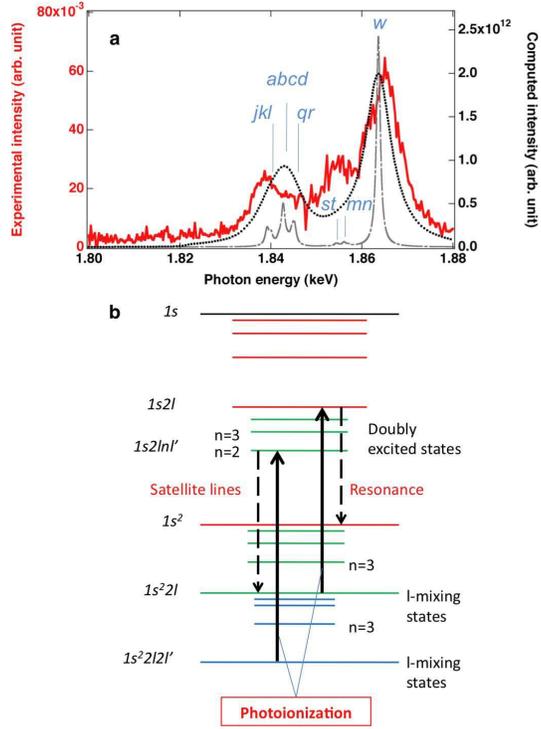}
	\end{center}
    \caption{(a) Comparison between the experimental (solid) and computed (dotted) x-ray spectra from the photoionized silicon plasma. Fine structure of computed spectrum (dot-line) is also drawn with key letters according to Gabriel's definition. (b) Energy diagram relevant to the photoionized plasma generated in laboratory. The central result of the present work is that the good agreement between the experimental and computational spectra was obtained when the model assumed that He-like $1s2p$ excited state is generated only by photoionization of a $K$-shell electron in a Li-like ground state ion ($1s^2 2\ell, \ell=0,1$). In a similar way, the excited Li-like ion $1s 2\ell 2\ell' (\ell,\ell'=0,1)$, which is the upper configuration for the satellite lines, is generated by photoionization of a K-shell electron in a Be-like ion $1s^2 2\ell 2\ell' (\ell,\ell' = 0,1$). \label{fig: detail spectrum}}
\end{figure}

	In order to clarify the physical processes that lead to the experimental spectrum, computational modeling was carried out.
	The computational modeling includes the simulation of the incident radiation field and its effects on the time-dependent energy and ionization state balance in the photoionized silicon plasma.
	The computations indicate that the maximum increment of electron temperature of 20 eV shortly after the x-ray irradiation peaks, and slowly cools down at later time.
	At such low temperatures, photoionization, radiative recombination and spontaneous decay of the excited states were identified as the dominant processes influencing the population of the ionization and excitation state distributions.
    Typical time scales of photoionization of $L$-shell and $K$-shell electrons from silicon ions are calculated to be 50 ps and 275 ps, respectively.
	Most of the electrons outer to the $K$-shell are, therefore, photo- or Auger-ionized during the radiation pulse (160 ps).
    The computed average ionization reaches a maximum of 11.9 shortly after the peak of the radiation pulse, later the recombination process gradually reduces it.
    The atomic energy levels and transition probabilities were obtained from the HULLAC code \cite{Bar-Shalom_97}.
    The computational spectrum is shown in Fig. \ref{fig: detail spectrum} (a) (dotted line).
	Line broadening was taken into account in the comparison.
    Fine structure of the computed spectrum (dashed-and-dotted line) is also drawn in Fig. \ref{fig: detail spectrum} (a) with key letters labeled according to Gabriel's convention \cite{Gabriel_72_MNRAS}.

      The computational spectrum exhibits two spectral peaks: (i) the $1s^2\ {}^1S_0$  - $1s2p\ {}^1P_1$ resonance transition in He-like silicon ions at 1.863 keV, and (ii) a combination of three satellite lines from Li-like ions around 1.840 keV, see Fig. \ref{fig: detail spectrum} (b). The computations indicate that the resonance line stems from photoionization of a $K$-shell electron in Li-like ions followed by radiative decay of an $L$-electron into the $K$-shell vacancy. The satellite lines originate in a similar mechanism in Be-like ions. This is a significant difference relative to collisional plasmas where such doubly excited states are generated, in general, by electron impact excitation or dielectronic recombination.
	The weak bump near 1.85 - 1.86 keV stems from the intercombination lines ($1s^2 \ {}^1S_0$ - $1s2p \ {}^3P_1$).

	We obtained x-ray spectra in the laboratory that resemble those observed astronomically, although their interpretations are contradicting.
	The spectral peak near 1.84 keV in Figs. \ref{fig: spectra} (b) and (c) is thought to be a forbidden line ($1s^2 \ {}^1S_0$ - $1s2s \ {}^3S_1$) of He-like silicon ions in astronomy \cite{Schulz_02_ApJ}.
		The HULLAC code indicates that the forbidden line of He-like silicon ions should occur at 1.84 keV. This photon energy is coincidentally almost equal to that of the satellite lines, but the spontaneous decay rate (Einstein $A$ coefficient) of the forbidden line is 10$^{-8}$ times that of the resonant line.
	Consequently the population of the $1s2s$ state must be 10$^8$ times larger than that of the $1s2p$ state, if the peaks near 1.84 eV do in fact stem from the forbidden transition.
	We have estimated the time-scales of several possible mechanisms that could generate an overpopulation of the $2s$ metastable state. All these time-scales turned out to be much longer than the lifetime of our photoionized plasma.
	It therefore seems difficult to believe that electrons accumulated in the $1s2s$ state in the present experiment.
	Nevertheless, it is still too early to make conclusions about the origin of the 1.84 keV peak in the astronomical case, because the radiation flux is pulsed in laboratory but continuous from compact objects.
	More experimental and computational studies are necessary to better understand this point.
    It is worthwhile to note that the Li-like satellite lines, which are generated from Be-like species by photoionization, should be incorporated also in the analysis of astronomical spectra.
	
The authors would like to acknowledge the dedicated technical support by the staff at the GEKKO-XII facility for the laser operation, target fabrication, and plasma diagnostics. We also acknowledge Prof. K. Masai of Tokyo Metropolitan University and Dr. T. Dotani of the Japanese Aerospace Exploration Agency, as well as their colleagues, for valuable discussions of photoionized plasma in astronomy. This research was partially supported by the Japanese Ministry of Education, Science, Sports, and Culture (MEXT),  Special Education and Research Expenses for "Laboratory Astrophysics with High-Power Laser" and Grant-in-Aid for Young Scientists (A) for "Non-LTE Photoionized Plasma Generation with Laser-Produced Blackbody Radiator (Grant No. 21684034)", and by the National Basic Research Program of China (973 Program, Grant No. 2007CB815102-03). 

\bibliographystyle{apsrev}
\bibliography{References}

\begin{thebibliography}{24}
\expandafter\ifx\csname natexlab\endcsname\relax\def\natexlab#1{#1}\fi
\expandafter\ifx\csname bibnamefont\endcsname\relax
  \def\bibnamefont#1{#1}\fi
\expandafter\ifx\csname bibfnamefont\endcsname\relax
  \def\bibfnamefont#1{#1}\fi
\expandafter\ifx\csname citenamefont\endcsname\relax
  \def\citenamefont#1{#1}\fi
\expandafter\ifx\csname url\endcsname\relax
  \def\url#1{\texttt{#1}}\fi
\expandafter\ifx\csname urlprefix\endcsname\relax\def\urlprefix{URL }\fi
\providecommand{\bibinfo}[2]{#2}
\providecommand{\eprint}[2][]{\url{#2}}

\bibitem[{\citenamefont{Tarter et~al.}(1969)\citenamefont{Tarter, Tucker, and
  Salpeter}}]{Tarter_69_ApJ}
\bibinfo{author}{\bibfnamefont{C.~B.} \bibnamefont{Tarter}},
  \bibinfo{author}{\bibfnamefont{W.~H.} \bibnamefont{Tucker}},
  \bibnamefont{and} \bibinfo{author}{\bibfnamefont{E.~E.}
  \bibnamefont{Salpeter}}, \bibinfo{journal}{Astrophys. J.}
  \textbf{\bibinfo{volume}{156}}, \bibinfo{pages}{943} (\bibinfo{year}{1969}).

\bibitem[{\citenamefont{Osterbrock}(1989)}]{Osterbrock_text}
\bibinfo{author}{\bibfnamefont{D.~E.} \bibnamefont{Osterbrock}},
  \emph{\bibinfo{title}{Astrophysics of Gaseous Nebulae and Active Galactic
  Nuclei}} (\bibinfo{publisher}{University Science Books},
  \bibinfo{address}{Mill Valley, CA}, \bibinfo{year}{1989}).

\bibitem[{\citenamefont{Liedahl}(1999)}]{Liedahl_99_text}
\bibinfo{author}{\bibfnamefont{D.~A.} \bibnamefont{Liedahl}},
  \emph{\bibinfo{title}{X-ray spectroscopy in astrophysics}}
  (\bibinfo{publisher}{Springer-Verlag}, \bibinfo{address}{Berlin},
  \bibinfo{year}{1999}), chap. \bibinfo{chapter}{The x-ray spectral properties
  of photoionized plasmas and transient plasmas}, p. \bibinfo{pages}{189}.

\bibitem[{\citenamefont{Ballantyne et~al.}(2002)\citenamefont{Ballantyne, Ross,
  and Fabian}}]{Ballantyne_02_MNRAS}
\bibinfo{author}{\bibfnamefont{D.~R.} \bibnamefont{Ballantyne}},
  \bibinfo{author}{\bibfnamefont{R.~R.} \bibnamefont{Ross}}, \bibnamefont{and}
  \bibinfo{author}{\bibfnamefont{A.~C.} \bibnamefont{Fabian}},
  \bibinfo{journal}{Mon. Not. R. Astron. Soc.} \textbf{\bibinfo{volume}{336}},
  \bibinfo{pages}{867} (\bibinfo{year}{2002}).

\bibitem[{\citenamefont{Nuckolls et~al.}(1972)\citenamefont{Nuckolls, Wood,
  Thiessen, and Zimmerman}}]{ICF_concept}
\bibinfo{author}{\bibfnamefont{J.}~\bibnamefont{Nuckolls}},
  \bibinfo{author}{\bibfnamefont{L.}~\bibnamefont{Wood}},
  \bibinfo{author}{\bibfnamefont{A.}~\bibnamefont{Thiessen}}, \bibnamefont{and}
  \bibinfo{author}{\bibfnamefont{G.~B.} \bibnamefont{Zimmerman}},
  \bibinfo{journal}{Nature} \textbf{\bibinfo{volume}{239}},
  \bibinfo{pages}{139} (\bibinfo{year}{1972}).

\bibitem[{\citenamefont{Paerels et~al.}(2000)\citenamefont{Paerels, Cottam,
  Sako, Liedahl, Brinkman, der Meer, Kaastra, and Predehl}}]{Paerels_00_ApJ}
\bibinfo{author}{\bibfnamefont{F.}~\bibnamefont{Paerels}},
  \bibinfo{author}{\bibfnamefont{J.}~\bibnamefont{Cottam}},
  \bibinfo{author}{\bibfnamefont{M.}~\bibnamefont{Sako}},
  \bibinfo{author}{\bibfnamefont{D.}~\bibnamefont{Liedahl}},
  \bibinfo{author}{\bibfnamefont{A.~C.} \bibnamefont{Brinkman}},
  \bibinfo{author}{\bibfnamefont{R.~L. J.~V.} \bibnamefont{der Meer}},
  \bibinfo{author}{\bibfnamefont{S.}~\bibnamefont{Kaastra}}, \bibnamefont{and}
  \bibinfo{author}{\bibfnamefont{P.}~\bibnamefont{Predehl}},
  \bibinfo{journal}{Astrophys. J. Lett.} \textbf{\bibinfo{volume}{533}},
  \bibinfo{pages}{L135} (\bibinfo{year}{2000}).

\bibitem[{\citenamefont{Liedahl and Paerels}(1996)}]{Liedahl_96_ApJ}
\bibinfo{author}{\bibfnamefont{D.~A.} \bibnamefont{Liedahl}} \bibnamefont{and}
  \bibinfo{author}{\bibfnamefont{F.}~\bibnamefont{Paerels}},
  \bibinfo{journal}{Astrophys. J.} \textbf{\bibinfo{volume}{468}},
  \bibinfo{pages}{133} (\bibinfo{year}{1996}).

\bibitem[{\citenamefont{Schulz et~al.}(2002)\citenamefont{Schulz, Canizares,
  Lee, and Sako}}]{Schulz_02_ApJ}
\bibinfo{author}{\bibfnamefont{N.~S.} \bibnamefont{Schulz}},
  \bibinfo{author}{\bibfnamefont{C.~R.} \bibnamefont{Canizares}},
  \bibinfo{author}{\bibfnamefont{J.~C.} \bibnamefont{Lee}}, \bibnamefont{and}
  \bibinfo{author}{\bibfnamefont{M.}~\bibnamefont{Sako}},
  \bibinfo{journal}{Astrophys. J. Lett.} \textbf{\bibinfo{volume}{564}},
  \bibinfo{pages}{L21} (\bibinfo{year}{2002}).

\bibitem[{\citenamefont{Goldstein et~al.}(2004)\citenamefont{Goldstein,
  Huenemoerder, and Blank}}]{Goldstein_04_AJ}
\bibinfo{author}{\bibfnamefont{G.}~\bibnamefont{Goldstein}},
  \bibinfo{author}{\bibfnamefont{D.~P.} \bibnamefont{Huenemoerder}},
  \bibnamefont{and} \bibinfo{author}{\bibfnamefont{D.}~\bibnamefont{Blank}},
  \bibinfo{journal}{Astronomical J.} \textbf{\bibinfo{volume}{127}},
  \bibinfo{pages}{2310} (\bibinfo{year}{2004}).

\bibitem[{\citenamefont{Watanabe et~al.}(2006)\citenamefont{Watanabe, Sako,
  Ishida, Ishisaki, Kahn, Kohmura, Nagase, Paerels, and
  Takahashi}}]{Watanabe_06_ApJ}
\bibinfo{author}{\bibfnamefont{S.}~\bibnamefont{Watanabe}},
  \bibinfo{author}{\bibfnamefont{M.}~\bibnamefont{Sako}},
  \bibinfo{author}{\bibfnamefont{M.}~\bibnamefont{Ishida}},
  \bibinfo{author}{\bibfnamefont{Y.}~\bibnamefont{Ishisaki}},
  \bibinfo{author}{\bibfnamefont{S.~M.} \bibnamefont{Kahn}},
  \bibinfo{author}{\bibfnamefont{T.}~\bibnamefont{Kohmura}},
  \bibinfo{author}{\bibfnamefont{F.}~\bibnamefont{Nagase}},
  \bibinfo{author}{\bibfnamefont{F.}~\bibnamefont{Paerels}}, \bibnamefont{and}
  \bibinfo{author}{\bibfnamefont{T.}~\bibnamefont{Takahashi}},
  \bibinfo{journal}{Astrophys. J.} \textbf{\bibinfo{volume}{651}},
  \bibinfo{pages}{421} (\bibinfo{year}{2006}).

\bibitem[{\citenamefont{Kallman et~al.}(1996)\citenamefont{Kallman, Liedahl,
  Osterheld, Goldstein, and Kahn}}]{Kallman_96_ApJ}
\bibinfo{author}{\bibfnamefont{T.~R.} \bibnamefont{Kallman}},
  \bibinfo{author}{\bibfnamefont{D.}~\bibnamefont{Liedahl}},
  \bibinfo{author}{\bibfnamefont{A.}~\bibnamefont{Osterheld}},
  \bibinfo{author}{\bibfnamefont{W.}~\bibnamefont{Goldstein}},
  \bibnamefont{and} \bibinfo{author}{\bibfnamefont{S.}~\bibnamefont{Kahn}},
  \bibinfo{journal}{Astrophys. J.} \textbf{\bibinfo{volume}{465}},
  \bibinfo{pages}{994} (\bibinfo{year}{1996}).

\bibitem[{\citenamefont{Ferland et~al.}(1998)\citenamefont{Ferland, Korista,
  Verner, Ferguson, Kingdon, and Verner}}]{Ferland_98_PASP}
\bibinfo{author}{\bibfnamefont{G.~J.} \bibnamefont{Ferland}},
  \bibinfo{author}{\bibfnamefont{K.~T.} \bibnamefont{Korista}},
  \bibinfo{author}{\bibfnamefont{D.~A.} \bibnamefont{Verner}},
  \bibinfo{author}{\bibfnamefont{J.~W.} \bibnamefont{Ferguson}},
  \bibinfo{author}{\bibfnamefont{J.~B.} \bibnamefont{Kingdon}},
  \bibnamefont{and} \bibinfo{author}{\bibfnamefont{E.~M.}
  \bibnamefont{Verner}}, \bibinfo{journal}{Pub. Astro. Soc. Pacific}
  \textbf{\bibinfo{volume}{110}}, \bibinfo{pages}{761} (\bibinfo{year}{1998}).

\bibitem[{\citenamefont{Rose}(1998)}]{Rose_98_JPB}
\bibinfo{author}{\bibfnamefont{S.~J.} \bibnamefont{Rose}}, \bibinfo{journal}{J.
  Phys. B} \textbf{\bibinfo{volume}{31}}, \bibinfo{pages}{2129}
  (\bibinfo{year}{1998}).

\bibitem[{\citenamefont{Chung et~al.}(2005)\citenamefont{Chung, Chen, Morgan,
  Ralchenko, and Lee}}]{Chung_05_HEDP}
\bibinfo{author}{\bibfnamefont{H.~K.} \bibnamefont{Chung}},
  \bibinfo{author}{\bibfnamefont{M.~H.} \bibnamefont{Chen}},
  \bibinfo{author}{\bibfnamefont{W.~L.} \bibnamefont{Morgan}},
  \bibinfo{author}{\bibfnamefont{Y.}~\bibnamefont{Ralchenko}},
  \bibnamefont{and} \bibinfo{author}{\bibfnamefont{R.~W.} \bibnamefont{Lee}},
  \bibinfo{journal}{High Energy Density Physics} \textbf{\bibinfo{volume}{1}},
  \bibinfo{pages}{3} (\bibinfo{year}{2005}).

\bibitem[{\citenamefont{Rose et~al.}(2004)\citenamefont{Rose, van Hoof,
  Jonauskas, Keenan, Kisielius, Ramsbottom, Foord, Heeter, and
  Springer}}]{Rose_04_JPB}
\bibinfo{author}{\bibfnamefont{S.~J.} \bibnamefont{Rose}},
  \bibinfo{author}{\bibfnamefont{P.~A.~M.} \bibnamefont{van Hoof}},
  \bibinfo{author}{\bibfnamefont{V.}~\bibnamefont{Jonauskas}},
  \bibinfo{author}{\bibfnamefont{F.~P.} \bibnamefont{Keenan}},
  \bibinfo{author}{\bibfnamefont{R.}~\bibnamefont{Kisielius}},
  \bibinfo{author}{\bibfnamefont{C.}~\bibnamefont{Ramsbottom}},
  \bibinfo{author}{\bibfnamefont{M.~E.} \bibnamefont{Foord}},
  \bibinfo{author}{\bibfnamefont{R.~F.} \bibnamefont{Heeter}},
  \bibnamefont{and} \bibinfo{author}{\bibfnamefont{P.~T.}
  \bibnamefont{Springer}}, \bibinfo{journal}{J. Phys. B}
  \textbf{\bibinfo{volume}{37}}, \bibinfo{pages}{L337} (\bibinfo{year}{2004}).

\bibitem[{\citenamefont{Remington et~al.}(1999)\citenamefont{Remington, Arnett,
  Drake, and Takabe}}]{Remington_99_Science}
\bibinfo{author}{\bibfnamefont{B.~A.} \bibnamefont{Remington}},
  \bibinfo{author}{\bibfnamefont{D.}~\bibnamefont{Arnett}},
  \bibinfo{author}{\bibfnamefont{R.~P.} \bibnamefont{Drake}}, \bibnamefont{and}
  \bibinfo{author}{\bibfnamefont{H.}~\bibnamefont{Takabe}},
  \bibinfo{journal}{Science} \textbf{\bibinfo{volume}{284}},
  \bibinfo{pages}{1488} (\bibinfo{year}{1999}).

\bibitem[{\citenamefont{Remington et~al.}(2006)\citenamefont{Remington, Drake,
  and Ryutov}}]{Remington_06_RMP}
\bibinfo{author}{\bibfnamefont{B.~A.} \bibnamefont{Remington}},
  \bibinfo{author}{\bibfnamefont{R.~P.} \bibnamefont{Drake}}, \bibnamefont{and}
  \bibinfo{author}{\bibfnamefont{D.~D.} \bibnamefont{Ryutov}},
  \bibinfo{journal}{Rev. Mod. Phys.} \textbf{\bibinfo{volume}{78}},
  \bibinfo{pages}{755} (\bibinfo{year}{2006}).

\bibitem[{\citenamefont{Foord et~al.}(2004)\citenamefont{Foord, Heeter, van
  Hoof, Thoe, Bailey, Cuneo, Chung, Liedahl, Fournier, Chandler
  et~al.}}]{Foord_04_PRL}
\bibinfo{author}{\bibfnamefont{M.~E.} \bibnamefont{Foord}},
  \bibinfo{author}{\bibfnamefont{R.~F.} \bibnamefont{Heeter}},
  \bibinfo{author}{\bibfnamefont{P.~A.~M.} \bibnamefont{van Hoof}},
  \bibinfo{author}{\bibfnamefont{R.~S.} \bibnamefont{Thoe}},
  \bibinfo{author}{\bibfnamefont{J.~E.} \bibnamefont{Bailey}},
  \bibinfo{author}{\bibfnamefont{M.~E.} \bibnamefont{Cuneo}},
  \bibinfo{author}{\bibfnamefont{H.~K.} \bibnamefont{Chung}},
  \bibinfo{author}{\bibfnamefont{D.~A.} \bibnamefont{Liedahl}},
  \bibinfo{author}{\bibfnamefont{K.~B.} \bibnamefont{Fournier}},
  \bibinfo{author}{\bibfnamefont{G.~A.} \bibnamefont{Chandler}},
  \bibnamefont{et~al.}, \bibinfo{journal}{Phys. Rev. Lett.}
  \textbf{\bibinfo{volume}{93}}, \bibinfo{pages}{055002}
  (\bibinfo{year}{2004}).

\bibitem[{\citenamefont{Wang et~al.}(2008)\citenamefont{Wang, Fujioka,
  Nishimura, Kato, Li, Zhao, Zhang, and Takabe}}]{Wang_08_PoP}
\bibinfo{author}{\bibfnamefont{F.}~\bibnamefont{Wang}},
  \bibinfo{author}{\bibfnamefont{S.}~\bibnamefont{Fujioka}},
  \bibinfo{author}{\bibfnamefont{H.}~\bibnamefont{Nishimura}},
  \bibinfo{author}{\bibfnamefont{D.}~\bibnamefont{Kato}},
  \bibinfo{author}{\bibfnamefont{Y.}~\bibnamefont{Li}},
  \bibinfo{author}{\bibfnamefont{G.}~\bibnamefont{Zhao}},
  \bibinfo{author}{\bibfnamefont{J.}~\bibnamefont{Zhang}}, \bibnamefont{and}
  \bibinfo{author}{\bibfnamefont{H.}~\bibnamefont{Takabe}},
  \bibinfo{journal}{Phys. Plasmas} \textbf{\bibinfo{volume}{15}},
  \bibinfo{pages}{073108} (\bibinfo{year}{2008}).

\bibitem[{\citenamefont{Giacconi et~al.}(1967)\citenamefont{Giacconi,
  Gorenstein, Gursky, and Waters}}]{Giacconi_67_APL}
\bibinfo{author}{\bibfnamefont{R.}~\bibnamefont{Giacconi}},
  \bibinfo{author}{\bibfnamefont{P.}~\bibnamefont{Gorenstein}},
  \bibinfo{author}{\bibfnamefont{H.}~\bibnamefont{Gursky}}, \bibnamefont{and}
  \bibinfo{author}{\bibfnamefont{J.~R.} \bibnamefont{Waters}},
  \bibinfo{journal}{Astrophys. J. Lett.} \textbf{\bibinfo{volume}{148}},
  \bibinfo{pages}{L119} (\bibinfo{year}{1967}).

\bibitem[{\citenamefont{Shakura and Sunyaev}(1973)}]{Shakura_73_AA}
\bibinfo{author}{\bibfnamefont{N.~I.} \bibnamefont{Shakura}} \bibnamefont{and}
  \bibinfo{author}{\bibfnamefont{R.~A.} \bibnamefont{Sunyaev}},
  \bibinfo{journal}{Astron. \& Astrophys.} \textbf{\bibinfo{volume}{24}},
  \bibinfo{pages}{337} (\bibinfo{year}{1973}).

\bibitem[{\citenamefont{Yamanaka et~al.}(1981)\citenamefont{Yamanaka, Kato,
  Izawa, Yoshida, Yamanaka, Sasaki, Nakatsuka, Mochizuki, Kuroda, and
  Nakai}}]{Yamanaka_81_IEEEQE}
\bibinfo{author}{\bibfnamefont{C.}~\bibnamefont{Yamanaka}},
  \bibinfo{author}{\bibfnamefont{Y.}~\bibnamefont{Kato}},
  \bibinfo{author}{\bibfnamefont{Y.}~\bibnamefont{Izawa}},
  \bibinfo{author}{\bibfnamefont{K.}~\bibnamefont{Yoshida}},
  \bibinfo{author}{\bibfnamefont{T.}~\bibnamefont{Yamanaka}},
  \bibinfo{author}{\bibfnamefont{T.}~\bibnamefont{Sasaki}},
  \bibinfo{author}{\bibfnamefont{M.}~\bibnamefont{Nakatsuka}},
  \bibinfo{author}{\bibfnamefont{T.}~\bibnamefont{Mochizuki}},
  \bibinfo{author}{\bibfnamefont{J.}~\bibnamefont{Kuroda}}, \bibnamefont{and}
  \bibinfo{author}{\bibfnamefont{S.}~\bibnamefont{Nakai}},
  \bibinfo{journal}{IEEE J. Quantum Electron.}
  \textbf{\bibinfo{volume}{QE-17}}, \bibinfo{pages}{1639}
  (\bibinfo{year}{1981}).

\bibitem[{\citenamefont{Bar-Shalom et~al.}(1997)\citenamefont{Bar-Shalom, Oreg,
  and Klapisch}}]{Bar-Shalom_97}
\bibinfo{author}{\bibfnamefont{A.}~\bibnamefont{Bar-Shalom}},
  \bibinfo{author}{\bibfnamefont{J.}~\bibnamefont{Oreg}}, \bibnamefont{and}
  \bibinfo{author}{\bibfnamefont{M.}~\bibnamefont{Klapisch}},
  \bibinfo{journal}{Phys. Rev. E} \textbf{\bibinfo{volume}{56}},
  \bibinfo{pages}{R70} (\bibinfo{year}{1997}).

\bibitem[{\citenamefont{Gabriel}(1972)}]{Gabriel_72_MNRAS}
\bibinfo{author}{\bibfnamefont{A.~H.} \bibnamefont{Gabriel}},
  \bibinfo{journal}{Mon. Not. R. Astr. Soc.} \textbf{\bibinfo{volume}{160}},
  \bibinfo{pages}{99} (\bibinfo{year}{1972}).

\end{thebibliography}

\end{document}